\documentclass[conference]{IEEEtran}
\usepackage{amsmath,amssymb,amsfonts}
\usepackage{algorithmic,microtype}
\usepackage{graphicx}
\usepackage{dsfont}
\usepackage{textcomp}
\usepackage{xcolor}
\usepackage{mathtools}
\usepackage{array}
\newcolumntype{L}[1]{>{\raggedright\let\newline\\\arraybackslash\hspace{0pt}}m{#1}}
\newcolumntype{C}[1]{>{\centering\let\newline\\\arraybackslash\hspace{0pt}}m{#1}}
\newcolumntype{R}[1]{>{\raggedleft\let\newline\\\arraybackslash\hspace{0pt}}m{#1}}

\def\BibTeX{{\rm B\kern-.05em{\sc i\kern-.025em b}\kern-.08em
    T\kern-.1667em\lower.7ex\hbox{E}\kern-.125emX}}

\begin{document}
\linespread{0.94}
\title{\huge Study of Multiuser Scheduling with Enhanced Greedy Techniques for Multicell and Cell-Free Massive MIMO Networks \vspace{-0.5em}}

\author{Saeed Mashdour$^{\star}$, Rodrigo C. de Lamare $^{\star,\dagger}$ and João P. S. H. Lima $^{\ddagger}$ \\  $^{\star}$ Centre for Telecommunications Studies, Pontifical Catholic University of Rio de Janeiro, Brazil \\
$^{\dagger}$ Department of Electronic Engineering, University of York, United Kingdom \\
$^{\ddagger}$ CPqD, Campinas, Brazil \\
smashdour@gmail.com, delamare@cetuc.puc-rio.br, jsales@cpqd.com.br  \vspace{-2em}\\
\thanks{This work was supported by CNPq and CPqD.}}

\maketitle

\begin{abstract}
In this work, we investigate the sum-rate performance of multicell and cell-free massive MIMO systems using linear precoding and multiuser scheduling algorithms. We consider the use of a network-centric clustering approach to reduce the computational complexity of the techniques applied to the cell-free system. We then develop a greedy algorithm that considers multiple candidates for the subset of users to be scheduled and that approaches the performance of the optimal exhaustive search. We assess the proposed and existing scheduling algorithms in both multicell and cell-free networks with the same coverage area. Numerical results illustrate the sum-rate performance of the proposed scheduling algorithm against existing approaches.\\
\end{abstract}

\begin{IEEEkeywords}
Massive MIMO, multiuser scheduling, multicell systems, cell-free systems, clustering. \vspace{-1em}
\end{IEEEkeywords}

\section{Introduction}

Multicell multiuser MIMO (MU-MIMO) systems have been widely used in wireless networks to improve the transmission rate, spectral and energy efficiencies using antenna arrays at each base station (BS), which can serve several user terminals simultaneously in each cell \cite{Multi-Objective, Energy, Truly}. In these systems, different beamforming or precoding techniques such as zero forcing (ZF) or minimum mean squared error (MMSE) \cite{mmimo,wence,rmmseprec,plsprec,gbd,wlbd,tds,spa,bbprec,siprec,memd,bfpeg,baplnc,mbthp,rmbthp,rsrbd,rsthp,lrcc,cfrprec,zcprec} are used to improve the system performance by maximizing the received signal power or by resorting interference cancellation at the receiver side \cite{Enhanced, Equal-rate, HybridMMSE}.

When there are many users in a massive MU-MIMO network user scheduling is an important tool to achieve a desirable sum capacity and spectral efficiency\cite{DistributedUser}. This is key especially if each user requires a high data rate. In addition, user scheduling is required when pilot contamination reduces the transmit power for each user
in Massive MIMO \cite{Scheduling, Noncooperative}. Furthermore, when the number of users is larger than the number of transmit antennas, user scheduling is fundamental to achieve a desirable sum capacity \cite{OnDownlink}. Accordingly, a great deal of  research has been done in the context of user scheduling in massive MIMO networks. In \cite{OptimizingUser}, three methods for user scheduling has been presented to optimize the ergodic sum-rate. A joint user scheduling and transceiver design scheme for cross-link interference suppression that works on per-frame basis in interfering massive MIMO multi-cell scenarios has been reported in \cite{JointUserScheduling}. In \cite{OnDownlink}, a user-selection method based on simple zero-forcing beamforming with selection (ZFS) has been proposed which  attains a significant fraction of sum capacity and throughput of optimal method.

Unlike multicell massive MIMO systems where users in each cell are served by a BS, a network with the same area of cells including randomly located single-antenna access points (APs) serving all users in the same time-frequency resource was introduced in \cite{CellFreeMassive, CellFree, Precoding} and called cell-free massive MIMO which presents higher throughput compared with small cells. Cell-free massive MIMO is compared with cellular massive MIMO systems in \cite{CellFreeversus}, and it is shown that cell-free systems can provide significantly higher spectral efficiency for all users. Since the information of all APs and users are required in cell-free massive MIMO, it is suggested in literature to serve each user by a subset of APs to limit data sharing and computational complexity via two network-centric and user-centric approaches \cite{Scalable}. The first approach divides the APs into several clusters, where each includes different APs from other clusters. The latter, considers the AP subset providing the best channel conditions for each user which could be dynamic.  

In this paper, we investigate the sum-rate performance of the multicell and cell-free massive MIMO systems using greedy user scheduling based on ZFS algorithm in the downlink \footnote{{This work was supported by CNPq and CPqD.}}. We consider the use of a network-centric clustering approach to reduce the computational complexity of the techniques applied to the cell-free systems \cite{cesg}. We then develop a greedy algorithm that considers multiple candidates for the subset of users to be scheduled and that approaches the performance of the optimal exhaustive search. We assess the proposed and existing scheduling algorithms in both multicell and cell-free networks with the same coverage area. Numerical results illustrate the sum-rate performance of the proposed scheduling algorithm against existing approaches.

{\it Notation}: Throughout the paper, $\left [ x \right ]_{+}=\textup{max}\left \{ 0,x \right \}$, $\mathbf{I}_{n}$ denotes the $n\times n$ identity matrix, the complex normal distribution is represented by $\mathcal{CN}\left ( .,. \right )$, superscripts $^{T}$ ,$^{\ast}$ , and $^{H}$ denote transpose,
complex conjugate and hermitian operations respectively, $A\cup B $ is union of sets $A$ and $B$, and $A\setminus B $ shows exclusion of set $B$ from set $A$.

\section{System Model}
We consider an area for both the multicell and cell-free networks so that we can have a fair comparison between these networks. There are $K$ single antenna users distributed in the whole area. The multicell network consists of $L$ cells each including $K_{c}=\frac{K}{L}$ users and a BS equipped with $N_{t}$ antennas. We also consider the cell-free system with $M=L\times N_{t}$ randomly located single antenna APs. 

\subsection{Multicell Channel and Signal Model}\label{AA}

We consider $\mathcal{L}=\left \{ 1,2,\cdots ,L \right \}$ as ths set of all existing cells in the area. We also model $h_{s_{mk}}$ as the Rayleigh fading coefficient between the $m$th transmit antenna and $k$th receive antenna in the cell $s$ and denote the row vector $\mathbf{h}_{s_{m}}=\left [ h_{s_{m1}}\quad h_{s_{m2}} \quad\cdots \quad h_{s_{mK_{c}}} \right ]$. Then, $\mathbf{H}_{l} \in \mathbb{C}^{K_{c}\times N_{t}}
$ channel matrix including channel coefficient from the BS $s$ to all users located in the related cell is
\begin{equation}
    \mathbf{H}_{l}=\left [ \mathbf{h}_{l_{1}}^{T}\quad\mathbf{h}_{l_{2}}^{T} \quad\cdots \quad \mathbf{h}_{l_{N_{t}}}^{T} \right ]
\end{equation}
Considering $\mathbf{\mathbf{x}}_{s}=\left [ x_{s_{1}}\quad \cdots \quad x_{s_{K_{c}}}\right ]^{T}$ as the vector
of information symbols from BS $s$ intended for all its related users with $\mathbf{x}_{s}\sim \mathcal{CN}\left ( \mathbf{0},\mathbf{I}_{K_{c}} \right )$, and $\mathbf{P}_{s}\in \mathbb{C}^{N_{T}\times K_{c}}$ as precoding matrix, the total downlink signal received by all users in cell $s$ is given by
\begin{equation}
\mathbf{y}_{s}=\mathbf{H}_{s}\mathbf{P}_{s}\mathbf{x}_{s}+\sum_{l=1,l\neq s}^{L}q_{sk}\mathbf{H}_{l}\mathbf{P}_{l}\mathbf{x}_{l}+\mathbf{w}_{s}
\end{equation}
where $\mathbf{w}_{s}=\left [ w_{s_{1}} \quad \cdots \quad w_{s_{K_{c}}} \right ]^{T}$ is the additive white Gaussian
noise vector of the users located in cell $s$ with $w_{s_{k}}\sim \mathcal{CN}\left ( 0,\sigma_{w}^{2} \right )$ and covariance matrix $\mathbf{C}_{\mathbf{w}_{s}}=\sigma _{w}^{2}\mathbf{I}_{K_{c}}$, $q_{sk}=\left [ \mathbf{Q} \right ]_{s,k}
$ and $\mathbf{Q}$ is the coupling matrix which describes the configuration of an interference model of a multiuser multicell system\cite{Distributed}. The term $\sum_{l=1,l\neq s}^{L}q_{sk}\mathbf{H}_{l}\mathbf{P}_{l}\mathbf{x}_{l}$ is considered as the interference caused by other cells called inter-cell interference (ICI). Since all noise vectors in each cell is independent of other cells, sum-rate of the received signals in the total network is achieved by summation of sum-rate of all the cells expressed by
\begin{equation} \label{eq:cap_mul}
    R_{t}=\sum_{s=1}^{L}R_{s}=\sum_{s=1}^{L}\log_{2}\left (\det \left [\mathbf{R}_{s}+\mathbf{I}_{K_{c}} \right ]\right),
\end{equation}
where the covariance matrix of the received signal is given by
\begin{equation}
     \mathbf{R}_{s}=\mathbf{H}_{s}\mathbf{P}_{s}\mathbf{P}_{s}^{H}\mathbf{H}_{s}^{H}\left (\sum_{l=1,l\neq s }^{L}\left | q_{sk} \right |^{2}\mathbf{H}_{l}\mathbf{P}_{l}\mathbf{P}_{l}^{H}\mathbf{H}_{l}^{H}+\sigma _{w}^{2}\mathbf{I}_{K_{c}}\right )^{-1}.
\end{equation}

\subsection{Cell-Free Channel and Signal Model}\label{BB}
According to \cite{CellFree, Precoding}, we use $g_{mk}=\sqrt{\beta _{mk}}h_{mk}$ to denote the cell-free channel coefficient between $m$th AP and $k$th user where $\beta _{mk}$ is the large-scale fading coefficient (path loss and shadowing
effects) and $ h_{mk}\sim \mathcal{CN}\left ( 0,1 \right )$ is the small-scale fading coefficient, defined
as independent and identically distributed (i.i.d) random variables (RVs)
that remain constant during a coherence interval and are independent over
different coherence intervals. Large scale coefficients are modeled as $\beta _{mk}=\textup{PL}_{mk}.10^{\frac{\sigma _{sh}z_{mk}}{10}}$ where $\textup{PL}_{mk}$ is the path loss and $10^{\frac{\sigma _{sh}z_{mk}}{10}}$
refers to the shadow fading with $\sigma _{sh}=8\textup{dB}$ and  $z_{mk}\sim\mathcal{N}\left ( 0,1 \right )$. According to \cite{Mobile}, the path loss is modeled as 

\begin{equation}
    \textup{PL}_{mk}=\left\{\begin{matrix}
-\textup{D}-35\log_{10}\left ( d_{mk} \right ), \textup{if }  d_{mk}>d_{1}& \\ 
 -\textup{D}-10\log_{10}\left ( d_{1}^{1.5} d_{mk}^2\right ), \textup{if }  d_{0}<d_{mk}\leq d_{1}& \\ 
 -\textup{D}-10\log_{10}\left ( d_{1}^{1.5} d_{0}^2 \right ), \textup{if }  d_{mk}\leq d_{0}&  
\end{matrix}\right.,
\end{equation}
where $d_{mk}$ denotes the distance between the $m$th AP and the
$k$th user,
\begin{equation}
\begin{multlined}
    \textup{D}=46.3+33.9\log_{10}\left ( f \right )-13.82\log_{10}\left ( h_{AP} \right ) \\
 -\left [ 1.11\log_{10}\left ( f \right )-0.7 \right ]h_{u}+1.56\log_{10}\left ( f \right )-0.8 
\end{multlined}
\end{equation}
where $f=1900$MHz is the carrier frequency, $h_{AP}=$15m and $h_{r}=$1.5m are the
AP and user antenna heights, respectively, $d_{0}=$10m and $d_{0}=$50m. When $d_{mk}\leq d_{1}$ there is no shadowing.

In downlink transmission, the signal received by the $k$th user is described by
\begin{equation}
    y_{k}=\sqrt{\rho _{f}}\mathbf{g}_{k}\mathbf{P}\mathbf{x}+w_{k}
\end{equation}
where $\rho _{f}$ is is maximum transmitted power of each antenna, $\mathbf{g}_{k}=\left [ g_{1k},\cdots g_{Mk} \right ]$ are the channel coefficients for user $k$, $\mathbf{P}\in \mathbb{C}^{M\times K}$ is the precoder matrix such as MMSE or ZF, $\mathbf{x}=\left [ x_{1},\cdots ,x_{K} \right ]^{T}$ is the zero mean symbol vector with $x_{k}$ the data symbol for user $k$ and $\mathbf{x}\sim \mathcal{CN}\left ( \mathbf{0},\mathbf{I}_{K} \right )$, and $w_{k}\sim \mathcal{CN}\left ( 0,\sigma_{w}^{2} \right )$ is the additive noise for user $k$. We consider elements of $\mathbf{s}$ mutually independent, and independent of all noise and
channel coefficients. Combining all the users, we have
\begin{equation}
    \mathbf{y}=\sqrt{\rho _{f}}\mathbf{G}^T\mathbf{P}\mathbf{x}+\mathbf{w}
\end{equation}
where $\mathbf{G}\in \mathbb{C}^{M\times K}$  is the channel matrix with elements $\left [ \mathbf{G} \right ]_{m,k}=g_{mk}$ and
$\mathbf{w}=\left [ w_{1},\cdots,w_{K}  \right ]^{T}$ is the noise vector. Considering noise covariane matrix as $\mathbf{C}_{\mathbf{w}}=\sigma _{w}^{2}\mathbf{I}_K$, the sum-rate of
the cell-free system can be computed by 
\begin{equation} \label{eq:cap_CF}
    R_{CF}=\log_{2}\left (  \det\left [\sigma _{w}^{-2}\rho _{f} \mathbf{G}^{T}\mathbf{P}\mathbf{P}^{H}\mathbf{G}^{\ast }+\mathbf{I}_K  \right ]\right )
\end{equation}

\section{ZFS User Scheduling and the Proposed Enhanced Scheduling Algorithm}

{Using an exhaustive search, we can schedule the user set with best performance among all possible user sets. However, it implies a high computational complexity which makes it impractical. Thus, alternative methods such as greedy algorithms are suggested in literature to reduce the selection complexity \cite{Lowcomplexity, LowcomplexityMIMO, Simplified}. They usually
consider a selection criterion and based on that a user with best match at each iteration is selected.}

In this section, we extend {the greedy} ZFS scheduling algorithm developed in \cite{OnDownlink} and to the scenarios of interest. Then, exploiting ZFS and introducing a strategy based on multiple candidates for choosing the subset of users, we develop an enhanced greedy algorithm which leads to a user subset closer to the optimal subset obtained by exhaustive search.

\subsection{ZFS Algorithm}\label{AA}

ZF precoder creates orthogonal channels between transmitter and receivers by inverting the channel matrix at the transmitter using precoding matrix $\mathbf{P}=\mathbf{H}^{H}\left ( \mathbf{H}\mathbf{H}^{H} \right )^{-1}$ where $\mathbf{H} \in \mathbb{C}^{K\times M}$ is the channel matrix. However, if $K> M$, $\mathbf{H}\mathbf{H}^{H}$ becomes singular and it is not possible to use ZF precoder. Therefore, it is required to schedule $n\leq M$ out of $K$ users as a set of users $S_{n}$ resulting in a row-reduced channel matrix $\mathbf{H}\left ( S_{n} \right )$ which gives the highest achievable sum-rate 
\begin{equation*}
\begin{aligned}
& \underset{1\leq n\leq M}{\text{max}}
& & \underset{S_{n}}{\text{max}}R_{zf}\left ( S_{n} \right ) \\
& \text{subject to}
& & \sum_{i \in S_{n}}\left [ \mu -\frac{1}{c_{i}\left ( S_{n} \right )} \right ]_{+}=P.
\end{aligned}
\end{equation*}
where $P$ is the upper limit of the signal covariance matrix $\textup{Trace}\left [ \mathbf{C}_{\mathbf{x}}\right ]\leq P$, $R_{zf}\left ( S_{n} \right )$ is throughput of ZF algorithm given by
\begin{equation}
    R_{zf}\left ( S_{n} \right )=\sum_{i \in S_{n}}\left [ \log_{2}\left ( \mu c_{i}\left ( S_{n} \right ) \right ) \right ]_{+}
\end{equation}
where $c_{i}\left ( S_{n} \right )=\left \{ \left [ \left ( \mathbf{H}\left ( S_{n} \right )\mathbf{H}\left ( S_{n} \right )^{H} \right )^{-1} \right ]_{ii} \right \}^{-1}$. Then, the reduced-complexity sub-optimal ZFS algorithm is outlined in the following pseudo code considering $\mathcal{K}=\left \{ 1,2,\cdots ,K \right \}$ as the set of indices of all $K$ users, $K_s$ as number of users to be scheduled, and $\mathbf{h}_{k}$ as the channel vector of user $k$. Note that we have used equal power loading to obtain $\mu$.
\begin{itemize}
    \item [1)] \textbf{Initialization} 
    \begin{itemize}
    \item set $n=1$
    \item find a user $s_{1}$ such that 
    
    $s_{1}=\underset{k\in \mathcal{K}}{\textup{argmax}}\mathbf{h}_{k}\mathbf{h}_{k}^{*}$
    
    \item set $S_{1}=\left \{ s_{1} \right \}$ and denote the achieved rate $R_{ZF}\left ( S_{1} \right )_{max}$
        \end{itemize}

    \item [2)] \textbf{while} $n<K_s$
    
    \begin{itemize}
    \item increase $n$ by 1
    \item find a user $s_{n}$ such that 
    
    $s_{n}=\underset{k\in\left (  \mathcal{K} \setminus  S_{n-1} \right )}{\textup{argmax}}R_{ZF}\left ( S_{n-1}\cup \left \{ k \right \} \right )$
    
    \item set $S_{n}=S_{n-1}\cup \left \{ s_{n} \right \}$ and denote the achieved rate $R_{ZF}\left ( S_{n} \right )_{max}$
    \item If $R_{ZF}\left ( S_{n} \right )_{max}\leq R_{ZF}\left ( S_{n-1} \right )_{max}$, breake and decrease $n$ by 1
    
        \end{itemize}
    \item [3)] \textbf{Precoding} $\mathbf{P}=\mathbf{H}\left ( S_{n} \right )^{*}\left (  \mathbf{H}\left ( S_{n} \right )\mathbf{H}\left ( S_{n} \right )^{*}\right )^{-1}$
     
     \end{itemize}
     
     \subsection{Enhanced Greedy Algorithm}\label{BB}
     
{Here}, we devise a scheduling strategy that assesses more sets of users so that we can achieve a system performance closer to the performance achieved by an exhaustive search while saving significant computational complexity. In this regard, we consider the  set achieved by ZFS as the first user set $S_{n\left ( 1 \right )}$. Then, we choose $k_{ex}$ as the least channel power user among the users of the first set which is called the first excluded user and is obtained by 
     \begin{equation}
         k_{ex\left (1\right )}=\underset{k\in S_{n\left ( 1 \right )}}{\textup{argmin}}\mathbf{h}_{k}\mathbf{h}_{k}^{*}
     \end{equation}
     We also select the user with the highest channel power from the remaining users other than the first selected set called first new user $k_{new}$ defined as
     \begin{equation}
         k_{new\left (1\right )}=\underset{k\in \mathcal{K}_{r\left ( 1\right )}}{\textup{argmax}}\mathbf{h}_{k}\mathbf{h}_{k}^{*}
     \end{equation}
     where $\mathcal{K}_{r\left ( 1\right )}=\mathcal{K}\setminus S_{n\left ( 1 \right )}$ is set of the remaining or unselected users. Substituting the excluded user by the new user in the first set, we achieve a new user set as the second set. Then, excluding the new user from the remaining users we achieve the second remaining user set. Repeating the described procedure for the second set and so on, we achieve $\frac{K-K_{s}}{2}$ sets together with the first set. Therefore, the user set $S_{n\left ( j \right )}$ and remaining user set $\mathcal{K}_{r\left ( j\right )}$, $j \in \left \{ 2,\cdots ,\frac{K-K_s}{2}+1 \right \}
$, are respectively derived as 
       
       \begin{equation}
        S_{n\left ( j \right )}=\left (S_{n\left ( j-1 \right )}\setminus k_{ex\left (j-1\right )} \right )\cup k_{new\left (j-1\right )}
     \end{equation}
     \begin{equation}
         \mathcal{K}_{r\left ( j\right )}=\mathcal{K}_{r\left ( j-1\right )}\setminus k_{new\left (j-1\right )}
      \end{equation}
      Thereafter, we assess all the considered sets to determine the best set $S_{n_{f}}$ using two different criteria each of which implying a different complexity to the system. The first criterion would be the sum-rates according to equations (\ref{eq:cap_mul}) and (\ref{eq:cap_CF}) for multicell and cell-free networks, respectively. The second one, would be $C_{s}\left ( S_{n\left ( i \right )} \right )$ the sum channel correlation among the users of $i$th set, $i \in \left \{ 1,2,\cdots ,\frac{K-K_s}{2}+1 \right \}$ defined as
     \begin{equation}
         C_s\left ( S_{n\left ( i \right )} \right )=\sum_{u \in S_{n\left ( i \right )}}\sum_{v \in S_{n\left ( i \right )},v\neq u}C_{u,v}
     \end{equation}
     where $C_{u,v}$ is channel correlation of the users $u$ and $v$ in the set $S_{n\left ( i \right )}$. Thus, depending on sum-rate or sum correlation criteria, the desired set is respectively derived as
     \begin{equation}
         S_{n_{f}}=\underset{S_{n} \in S_{n\left ( i \right )}}{\textup{argmax}}\left \{  R_{CF}\left ( S_{n} \right ) \right \}
     \end{equation}
    or alternatively as 
     \begin{equation}
         S_{n_{f}}=\underset{S_{n} \in S_{n\left ( i \right )}}{\textup{argmin}}\left \{   C_{s} \left ( S_{n} \right ) \right \}
     \end{equation}
      Accordingly, the enhanced greedy algorithm for cell-free network is outlined as follows. Note that for the multicell network, the number of users to be selected will change to $K_{c}$ and $K_{c_{s}}$ respectively, $R_{CF}$ is changed to $R_{t}$, and overall sum channel correlation is obtained by summation over all cells.
     \begin{itemize}
\bigskip

\item  $\textup{stage}=1$
    \item Finding initial user set $S_{n\left ( \textup{stage} \right )}$ using ZFS
    \item \textup{Compute}: $ R_{CF}\left ( S_{n\left ( \textup{stage} \right )}\right )$ or $ C_{s}\left ( S_{n\left ( \textup{stage} \right )} \right )$\% depending on the used criterion
    \item $\mathcal{K}_{r\left ( \textup{stage}\right )}=\mathcal{K}\setminus S_{n\left ( \textup{stage} \right )}$ \%  set of $K-K_s$ unselected users  
    \item $ k_{ex\left (\textup{stage}\right )}=\underset{k\in S_{n\left ( \textup{stage} \right )}}{\textup{argmin}}\mathbf{h}_{k}\mathbf{h}_{k}^{*}$ 
            
            \item  $k_{new\left (\textup{stage}\right )}=\underset{k\in \mathcal{K}_{r\left ( \textup{stage}\right )}}{\textup{argmax}}\mathbf{h}_{k}\mathbf{h}_{k}^{*}$
    
    \item for $\textup{stage}=2$ to $\frac{K-K_s}{2}+1$ \% we are considering half of the remaining users
        \begin{itemize}
            
             \item $S_{n\left ( \textup{stage} \right )}=\left (S_{n\left ( \textup{stage}-1 \right )}\setminus k_{ex\left (\textup{stage}-1\right )}  \right )\cup k_{new\left (\textup{stage}-1\right )}$ 
             
            \item $\mathcal{K}_{r\left ( \textup{stage}\right )}=\mathcal{K}_{r\left ( \textup{stage}-1\right )}\setminus k_{new\left (\textup{stage}-1\right )}$ 
          
          \item $ k_{ex\left (\textup{stage}\right )}=\underset{k\in S_{n\left ( \textup{stage} \right )}}{\textup{argmin}}\mathbf{h}_{k}\mathbf{h}_{k}^{*}$ 
            
            \item  $k_{new\left (\textup{stage}\right )}=\underset{k\in \mathcal{K}_{r\left ( \textup{stage}\right )}}{\textup{argmax}}\mathbf{h}_{k}\mathbf{h}_{k}^{*}$ 
          
          \item \textup{Compute}: $R_{CF}\left ( S_{n\left ( \textup{stage} \right )} \right )$ or $ C_{s}\left ( S_{n\left ( \textup{stage} \right )} \right )$
        \end{itemize}
        
        \item end for
        \item $S_{n_{f}}=\underset{S_{n} \in S_{n\left ( i \right )}}{\textup{argmax}}\left \{  R_{CF}\left ( S_{n} \right ) \right \}$  or $\underset{S_{n} \in S_{n\left ( i \right )}}{\textup{argmin}}\left \{   C_{s} \left ( S_{n} \right ) \right \}$
        
        \item Precoding
    \end{itemize}

\section{Simulation Results}
In this section, we assess in terms of sum-rates the proposed and existing scheduling algorithm in multicell and cell-free scenarios using Matlab. Note that CF and CoMP in the results represent cell-free and multicell systems, respectively. In particular, we consider a squared area of size 400m as the whole area of the cell-free system with $M=64$ single-antenna randomly located APs and $K=16$ uniformly distributed single antenna users. For the multicell system, we consider the same area divided in $L=4$ non-overlapping cells each including $K_{c}=\frac{K}{L}=4$ users with the same location as the cell-free system and a BS located at the center of the cell with $N_{t}=\frac{M}{L}=16$ antennas. {We adopt the cell-free channel model in \cite{CellFree, Precoding} and the cellular channel model in \ref{AA}, and consider a static channel over each transmission packet in both networks. In Table \ref{table:charachteristics}, we provide more details on the parameters used for simulation.}

\begin{table}[htb!]
\begin{small}
\caption{Characteristics of the Simulated Systems}
\begin{center}
\begin{tabular}{| L{2cm} | C{2cm} | R{2cm} |}
\hline
 &Multicell & Cell-free  \\ [0.5ex] 
 \hline
\hline
Carrier frequency& 1900MHz& 1900MHz \\
\hline
Symbol energy & $P_s$=1& $P_s$=1 \\
\hline 
Transmit power & $N_{t}\times P_{s}$& $M\times P_{s}$ \\
\hline
\end{tabular}
\label{table:charachteristics}
\end{center}
\end{small}
\end{table}

\begin{figure}
	\centering
		\includegraphics[width=.8\linewidth]{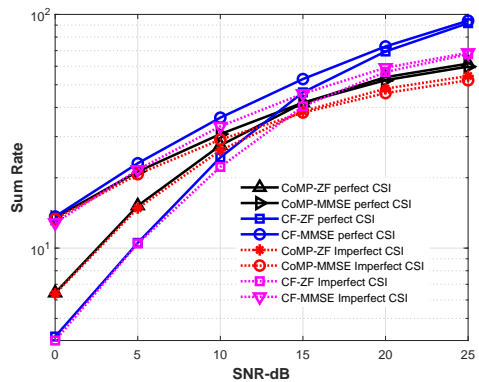}
		\vspace{-0.75em}
	\caption{\small{Performance comparison of cell-free and multicell networks considering all users}}
	\label{fig:CF and CoMP comparison}
\end{figure}

We have used ZF and MMSE precoders for performance comparison of multicell and cell-free networks and the corresponding sum-rates are shown in Fig. \ref{fig:CF and CoMP comparison} against the signal-to-noise ratio (SNR) when user scheduling is not considered. For all plots, the sum-rates increase with the SNR, however, the MMSE precoder has resulted in higher sum-rates as compared with the ZF precoder in both networks. The performance of the cell-free system has shown a significant superiority over the multicell system, which is mainly because cell-free network is not so much affected by the ICI near the cell borders which impairs the multicell system.

\begin{figure}
	\centering
		\includegraphics[width=.8\linewidth]{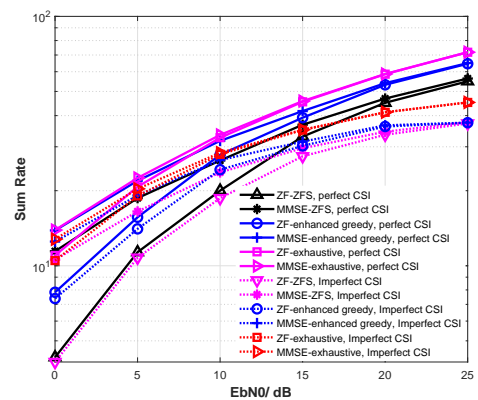}
		\vspace{-0.75em}
	\caption{\small{Performance of cell-free network with different scheduling methods}}
	\label{fig:CF with scheduling}
\end{figure}

\begin{figure}
	\centering
		\includegraphics[width=.8\linewidth]{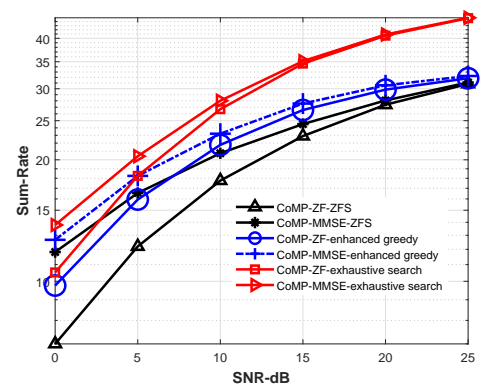}
		\vspace{-0.75em}
	\caption{\small{Performance of multicell network with different scheduling methods}}
	\label{fig:CoMP with scheduling}
\end{figure}

\begin{figure}
	\centering
		\includegraphics[width=.8\linewidth]{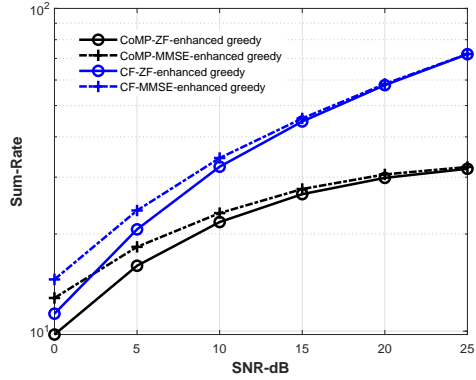}
		\vspace{-0.75em}
	\caption{\small{Performance comparison of cell-free and multicell networks using the proposed enhanced greedy method}}
	\label{fig:enhanced comparison}
\end{figure}

Figs. \ref{fig:CF with scheduling} and \ref{fig:CoMP with scheduling} show how different scheduling methods work in cell-free and multicell networks, respectively, when the sum-rate criterion is used. We have compared the performance of the systems implementing an exhaustive search, the proposed enhanced greedy and ZFS user scheduling algorithms to schedule half of the users when ZF and MMSE precoders are used. According to these figures, the proposed enhanced greedy method achieves an impressive improvement compared with ZFS and its performance is very close to the optimal exhaustive search method in cell-free networks. For multicell networks we also have a significant improvement. The performance of cell-free and multicell networks using the proposed enhanced greedy algorithm is also shown in Fig. \ref{fig:enhanced comparison}, where the cell-free system has outperformed the multicell system, as expected. 

\begin{figure}
	\centering
		\includegraphics[width=.8\linewidth]{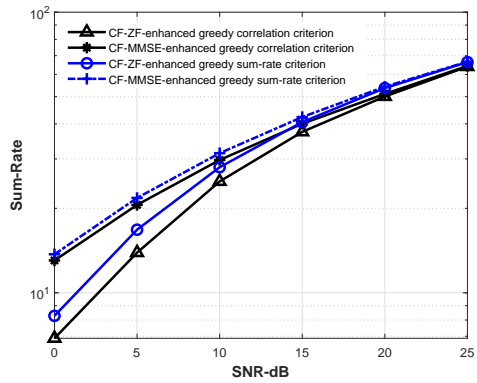}
		\vspace{-0.75em}
	\caption{\small{Performance of the enhanced greedy algorithm in a cell-free network for sum-rate and channel-correlation criteria}}
	\label{fig:criterion comparison}
\end{figure}

 In Fig. \ref{fig:criterion comparison}, the performance of the enhanced greedy algorithm in a cell-free network is shown when the sum-rate and channel-correlation criteria are used. We can see that using the sum-rate criterion provide an improvement in system performance while the results for the channel-correlation criterion are slighltly worse. The choice of the sum-rate and the channel-correlation criteria depends on the available computational power and application.

In Table \ref{table:complexity}, the computational complexity of the ZFS and the proposed enhanced greedy methods for the sum-rate and channel-correlation criteria are shown in cellular and cell-free networks with network-centring clustering. The results show that the improvement in the method using the sum-rate criterion comes at the cost of more required flops.

\begin{table}[htb!]
\begin{small}
\caption{Computational Complexity}
\begin{center}
\begin{tabular}{| L{2.5cm} | C{2.5cm} | R{2.5cm} |}
\hline
 &Multicell flops & Cell-free flops  \\ [0.5ex] 
 \hline
\hline
ZFS& 3580& 3580 \\
\hline
Enhanced greedy (channel-correlation) & 9484& 9484 \\
\hline
Enhanced greedy (sum capacity) & 12148& 12148 \\
\hline
\end{tabular}
\label{table:complexity}
\end{center}
\end{small}
\end{table}

\section{Conclusion}
We have proposed an enhanced greedy algorithm that extends the search procedure to find the user set with the best performance in cell-free and multicell MIMO systems. Numerical results show that there is a significant performance improvement in both cell-free and multicell systems using the proposed algorithm while linear MMSE and ZF precoders were used. The sum-rate criterion in the proposed enhanced greedy algorithm has outperformed the channel-correlation criterion at the cost of a slight increase in computational complexity.

\bibliographystyle{IEEEbib}
\bibliography{refs}

\end{document}